\documentclass[runningheads, envcountsame, a4paper]{llncs}

\usepackage[pdftex]{graphicx}
\usepackage{epstopdf}

\usepackage[hyphens]{url}
\usepackage{microtype}
\usepackage{enumitem}
\usepackage{amsmath}
\usepackage{amssymb}
\usepackage[mathscr]{euscript}
\usepackage{tikz}

\newcommand*\circled[1]{\tikz[baseline=(char.base)]{
            \node[shape=circle,fill,inner sep=1pt] (char) {\textcolor{white}{#1}};}}

\newtheorem{prop}{Proposition}
       
\usepackage{booktabs}
\usepackage{multirow}
\usepackage{subcaption}
\usepackage[misc]{ifsym}

\usepackage[linesnumbered, ruled]{algorithm2e}

\begin{document}

\title{Topological Anomaly Detection in Dynamic Multilayer Blockchain Networks}

\author{Ofori-Boateng, D. \inst{1}\Letter  \and Segovia Dominguez, I.\inst{2} \and Akcora, C.\inst{3}\and Kantarcioglu, M. \and Gel, Y.R.\inst{2}}

\titlerunning{Topological Anomaly Detection}

\authorrunning{Ofori-Boateng, D. et al.}

\institute{Portland State University, USA; \email{dorcas.oforiboateng@pdx.edu} \and 
University of Texas at Dallas, USA 
\and
University of Manitoba, Canada}

\maketitle

\begin{abstract}
Motivated by the recent surge of criminal activities with cross-cryptocurrency trades, we introduce a new topological perspective to structural anomaly detection in dynamic multilayer networks.
We postulate that anomalies in the underlying blockchain transaction graph that are composed of multiple layers are likely to also be manifested in anomalous patterns of the network shape properties. As such, we invoke the machinery of clique persistent homology on graphs to systematically and efficiently track evolution of the network shape and, as a result, to detect changes in the underlying network topology and geometry. We develop a new persistence summary for multilayer networks, called stacked persistence diagram, and prove its stability under input data perturbations.
We validate our new topological anomaly detection framework in application to dynamic multilayer networks from the Ethereum Blockchain and the Ripple Credit Network, and demonstrate that our stacked PD approach substantially outperforms state-of-art techniques.

\end{abstract}

\section{Introduction}
\label{Sec:Intro}

Due to the recent spike in popularity of crypto assets,  detecting anomalies in time evolving blockchain transaction networks has gained a new momentum. Here anomaly detection in dynamic graphs can be broadly defined as the problem of identifying instances within a sequence of graph observations where changes occur in the underlying structure of the graph. Indeed, these anomalies have significant implications, ranging from emergence of new ransomware (e.g., collecting ransom via cryptocurrencies) 
to financial manipulation. For example, in blockchain transaction networks, e.g., Ethereum, more frequent than expected appearance of particular subgraphs may indicate newly emerging malware or price pump-and-dump trading~\cite{xu2019anatomy}. Similarly, as recently shown by~\cite{weber2019anti}, the flow of coins on the Bitcoin graph provides important insights into money laundering schemes. As criminal, fraudulent, and illicit activities on blockchains continue to rise, with already stolen \$1.4B only in 2020, cryptocurrency criminals increasingly employ cross-cryptocurrency trades to hide their identity~\cite{Coindesk06022020}. As such, \cite{yousaf2019tracing} have recently shown that the analysis of links across multiple blockchain transaction graphs is critical for identifying emerging criminal and illicit activities on blockchain. However, while there exists a plethora of methods for network anomaly detection in single layer networks~\cite{ranshous2015anomaly,fernandes2019comprehensive,pourhabibi2020fraud}, there is yet {\it no single} method designed to detect anomalies in dynamic multilayer networks.

\vspace{0.2em}
\noindent{\bf Why TDA?} Motivated by the problem of tracking financial crime on blockchains, we develop a state-of-the-art methodology for anomaly detection on multilayer networks using Topological Data Analysis (TDA). Since crime on blockchains such as money laundering tends to involve multiple parties who possibly move funds across multiple cryptocurrency ledgers, one of our primary goals is to identify anomalous patterns in higher order graph connectivity. We postulate that anomalous higher order patterns can be detected using geometric and topological inference on graphs, that is, via a systematic analysis of the graph shape. To explore latent graph shape, we invoke the TDA machinery of the clique persistent homology (PH). PH allows to systematically infer qualitative and quantitative multi-lens geometric and topological structures from data directly and, hence, to enhance our understanding on the hidden role of geometry and topology in the system organization~\cite{Carlsson,TDAintro,WassermanTDA}. As a result, it may be intuitive to hypothesize that there shall be an intrinsic linkage between changes in the underlying graph structure and changes in the network shape which are then reflected in the extracted network topological characteristics. However, to the best of our knowledge, this paper is the first attempt to introduce TDA to anomaly detection in dynamic multilayer networks.

\vspace{0.2em}
\noindent{\bf Why Ethereum and Ripple?} Using the Blockchain global events timeline~\cite{Bitcoin_global_events}, we validate our methodology in application to anomaly detection in two multilayer blockchain network types, Ethereum and Ripple. While cryptocurrencies have already been adopted in payments, the recent surge in financial blockchain activity is largely due to platforms, such as Ethereum, which have brought algorithmic trading of digital assets by using Smart Contracts (i.e. short software code on the blockchain) in what is called Decentralized Finance~\cite{chen2020blockchain}. Assets include cryptocurrencies and crypto tokens as well. Hence, a given address (i.e. a node) may participate in transactions of multiple  digital assets. Looking at an individual asset transaction network alone (i.e. a single layer of the transaction graph) may provide a limited view. As a result, we need to consider multiple layers (e.g., a layer for each crypto token) and their interactions to detect anomalies. Resulting multilayer networks and participant activities are temporal, nuanced in the traded assets (e.g., coins, or fiat currencies), rich in network patterns and encode a new wave of financial heart-beat.  
The Ripple Credit Network transactions also comprise cross-border remittance transfers and even fiat currency trades,  allowing trading Ether, Bitcoin and other currencies on its system.

Our contributions, both in application and theory, are as follows:
\begin{itemize}[leftmargin=*]
    \item[1]  To the best of our knowledge, this is the first paper on anomaly detection in dynamic multilayer networks. 
    
    \item[2]  Our new methodology is based on the notion of clique persistent homology. To quantify topology of multilayer graphs,
    we introduce a multidimensional multi-set object, called the {\em stacked persistence diagram} (SPD). We prove that SPD is robust against minor input data perturbations w.r.t. bottleneck distance. 
    
   \item[3]  In the absence of the state-of-the-art anomaly detection methods for dynamic multilayer networks, we benchmark our topological anomaly detection (\texttt{TAD}) tool
   against a multiple testing framework, based on the strongest state-of-the-art (SOTA) methods for anomaly detection in single layer networks. 
  To control for family wise error rate (FWER) in the multiple testing framework, we use Bonferroni correction. We show that \texttt{TAD} substantially outperforms all competitors based on SOTA single layer solutions and the additional technique based on graph embedding.

    \item[4] We demonstrate utility of \texttt{TAD} 
    on Ethereum and Ripple blockchains, where digital assets worth billions of US Dollars are traded daily. We provide  Blockchain benchmark data  for anomaly detection on multilayer networks which is the first benchmark multilayer network dataset with ground-truth events, thereby further bridging AI with crypto-finance.

 \end{itemize}

\section{Related Work}
\label{Sec:Related_work}

{\bf Graph-Based Anomaly Detection:} Over recent years, there has been an increase in application of anomaly detection techniques for single layer graphs in interdisciplinary studies~\cite{yu2018netwalk,eswaran2018spotlight}. For example, \cite{Koutra2016deltacon} employed a graph-based measure (\texttt{DELTACON}) to assess connectivity between two graph structures with homogeneous node/edge attribution, and identified anomalous nodes/edges in the sequence of dynamic networks based on similarity deviations. With \texttt{DELTACON}, an event is flagged as anomalous if its similarity score lies below a threshold. In turn, \cite{Wang2017fast} devised a likelihood maximization tool that extracts a "feature" vector from individual networks, and uses dissimilarity between successive networks snapshots to classify anomalous or normal/regular events.  Procedure of \cite{Zhu2018hyper} segments network snapshots into separate clusters, infers local and global structure from individual nodes and their distribution via community detection and chronological ordering of the results in an effort to single-out potential anomalies. An online algorithm for detecting abrupt edge weight and structural changes in dynamic graphs has been recently introduced by \cite{Yoon2019fast}, but the method requires a pre-training data set to identify tuning parameters. In turn, \cite{mittal2018anomaly,suarez2018case,bansal2020ranking} discuss detection of malicious nodes in multiplex/multilayer networks. {Finally, \cite{dong2020modeling} proposed a score test for change point detection in multilayer networks that follow a multilayer weighted stochastic block model (SBM). However, the SBM assumption is infeasible for financial networks.} To our knowledge (see also the reviews by~~\cite{fernandes2019comprehensive,pourhabibi2020fraud}), \textit{there is no existing anomaly detection method designed for dynamic multilayer networks}.
\smallskip

\noindent\textbf{Blockchain:} Blockchain graphs have been extracted and analyzed for price prediction~\cite{greaves2015using,akcora2018bitcoin,kurbucz2019predicting}, measurement studies~\cite{victor2019measuring,lee2020measurements} and e-crime detection~\cite{chen2018detecting,akcora2019bitcoinheist}. Graph anomalies have been tracked to locate coins used in illegal activities, such as money laundering and blackmailing \cite{phetsouvanh2018egret}. These findings are known as taint analysis~\cite{di2015bitconeview}. Typically, a set of features are extracted from the blockchain graph and used in Machine Learning (ML) tasks. Here we bypass such a feature engineering step in learning on Blockchain networks.  
Ethereum structure has been analyzed by~\cite{ferretti2019ethereum,lee2020measurements}, while anomalies in Ethereum token prices have been evaluated using TDA tools~\cite{li2020dissecting}. In turn, Ripple has been assessed for its privacy aspects~\cite{moreno2016listening} and for health of the credit network~\cite{moreno2018mind}. However, multilayer analysis of blockchains have not been studied before.

\noindent {\bf TDA:} 
{Multiple recent papers show utility of TDA for developing early warning signals for crashes in the cryptocurrency market~\cite{gidea2018topological}, cryptocurrency price analytics~\cite{li2020dissecting}, and ransomware detection on blockchain transaction graphs~\cite{akcora2019bitcoinheist}. While TDA (as any other tool) cannot be viewed as a universal solution, TDA allows us to assess graph properties which are invariant under continuous deformations; hence it is likely to be one of the most robust tools for blockchain data analytics~\cite{zhao2020algebraic}}. TDA has been employed for visual detection of change points in single layer graphs~\cite{hajij2018visual}. 
In the multilayer network context, TDA has been used primarily for centrality ranking~\cite{taylor2019tunable}, including analysis of connectivity in the multiplex banking networks~\cite{de2017multiplex}, and clustering~\cite{yuvaraj2021topological}. Application of TDA to anomaly detection in multilayer networks is yet an unexplored area.

\noindent{\bf Multilayer Network Benchmark Data:} Multilayer networks receive an increasing attention in the last few years, due to their flexibility of modeling interconnected systems~\cite{aleta2019multilayer}.
There also exist several data repositories with multilayer graphs, e.g.~\cite{comunelab,TradeNetwork:Alves:2019}, but neither of them have publicly available benchmark data on multilayer graphs with ground truth for anomaly detection.

\section{The Mechanism of Persistent Homology}
\label{Sec:Preliminares}
Topology is the study of shapes. TDA and, in particular, {\em persistent homology} (PH) provides systematic mathematical means to extract the intrinsic shape properties of the observed data $\mathcal{X}$ (in our case $\mathcal{X}$ is a multilayer graph but $\mathcal{X}$ can be a point cloud in Euclidean or any finite metric space) that are invariant under continuous transformations. The key postulate is that $\mathcal{X}$ are sampled from some metric space $\mathcal{M}$ whose properties are lost due to sampling. The goal of PH is then to reconstruct the unknown topological and geometric structure of $\mathcal{M}$, based on systematic shape analysis of $\mathcal{X}$. In this paper, we introduce the PH concepts to analysis
of dynamic multilayer networks, starting by providing background on PH on graphs.

\begin{definition} Let $\mathcal{G} = (V, E, \omega)$ be a (weighted) graph, with vertex set $V$, edge set $E = \{e_{1},e_{2},\ldots\} \subseteq V\times V$, edge weights  $\omega= \omega(e): E \rightarrow{\mathbb{Z}^{+}}$ for all $e\in E$. 
\end{definition}

 At the initial stages of PH, we select a certain threshold $\nu_{*}>0$, and then we generate a subgraph $\mathcal{G}_{*} = (V, E_{*}, \omega_{*})$, such that $E_{*} =\{e \mid \omega(e)\leq\nu_{*}\}$, and $\omega_{*}(e) = \omega(e)$, for all $e \in E_{*}$. Then the observed graph $\mathcal{G}_{*}$ is equipped with a basic combinatorial object known as an {\em abstract simplicial complex}. Formally, a simplicial complex is defined as a collection $\mathcal{C}$ of finite subsets of $V(\mathcal{G})$ such that if $\sigma\in\mathcal{C}$ then $\tau\in\mathcal{C}$ for all $\tau\subseteq\sigma$. The basic unit of simplicial complexes is called the {\em simplex}, and if $|\sigma|=m+1$ then $\sigma$ is called an $m$-simplex. Specific to our analysis, we use a simplicial complex type called the {\em clique complex} to systematically and efficiently extract topological features from the observed $\mathcal{G}$. A clique complex $\mathscr{C}(\mathcal{G}_{*})$ is a simplicial complex with a simplex for every clique (i.e., a set of vertices of $\mathcal{G}_{*}$ such that any two points in the clique are adjacent) in $\mathcal{G}_{*}$. Furthermore, a $k$-clique community is formed whenever two $k$-cliques share $k-1$ vertices ($k\in \mathbb{Z}^{+}$). With a range of thresholds $\nu_1<\ldots<\nu_n$, we can obtain a hierarchically nested sequence of graphs $\mathcal{G}_{1} \subseteq \ldots \subseteq \mathcal{G}_{n}$ for any graph $\mathcal{G}$, where each individual subgraph will generate its own clique complex. Subsequently, the procedure which generates complexes from the nested sequence $\mathcal{G}_{1} \subseteq \ldots \subseteq \mathcal{G}_{n}$ is known as the {\em network filtration}, and the resultant complex generated by $\mathcal{G}$ is called a {\em filtered complex}~\cite{Zomorodian2010fast}. Particular to cliques, we construct clique complexes and then obtain the {\em clique filtration} $\mathscr{C}(\mathcal{G}_{1}) \subseteq \ldots \subseteq \mathscr{C}(\mathcal{G}_{n})$.

The mechanism of {\em clique persistent homology} involves tracking clique complexes over the filtration and quantifying lifespan of topological features/shapes such as loops, holes, and voids that appear and disappear at various thresholds $\nu_{*}$~\cite{zomorodian2010tidy,Rieck2017clique}.
We say that a topological feature is born at the $i$-th filtration step if it appears in $\mathscr{C}(\mathcal{G}_{i})$, and the topological feature dies at the $j$-th filtration step, if it disappears at $\mathscr{C}(\mathcal{G}_{j})$.
Hence, the lifespan of a topological feature is $\nu_{j}-\nu_{i}$.
The primary objective of TDA will then be to assess which topological features/shapes persist (i.e. have longer lifespan) over the clique filtration and, hence, are likelier to contain important structural information on the graph, and which topological features have shorter lifespan. The latter features are typically referred to as topological noise.  

One of the most widely used topological summaries is the {\em persistence diagram} (PD)~\cite{Carlsson,Zomorodian2010fast}. The PD is a  collection of points $(v_i,v_j) \in \mathbb{R}^{2}$ with each point corresponding to a topological feature, and the $x$- and $y$-coordinates representing birth and death times for the topological feature. Similarity between any two PDs, $D_a$ and $D_b$, can be computed using the Wasserstein ($W_r$) or the Bottleneck distances ($W_{\infty}$):
$$ {W_{r}(D_{a}, D_{b})=
 {\bigl(\inf_{\eta}\sum_{x\in {D_{a}}}\|x-\eta(x)\|^r_{\infty}\bigr)}^{1/r}}, 
 \quad
   W_{\infty}(D_{a},D_{b}) = \inf_{\eta}  \sup_{x\in {D_{a}}} \|x-\eta(x)\|_{\infty}.$$

\noindent Here $r\geq 1$, $\eta$ ranges over all bijections from $D_{a}\cup \Delta$ to $D_{b}\cup \Delta$, counting multiplicities, with $\Delta= \{(x, x) | x \in \mathbb{R}\}$
and $||z||_{\infty}=\max_i|z_i|$~\cite{Kerber_et_al2017,WassermanTDA}. {We evaluate both distances in the methodological development of the \texttt{TAD}.}

\begin{table}[ht]
 \caption{Description of Symbols Used}
  \label{tab:notations-tad}
    \centering
  \begingroup
  \begin{tabular}{lll}
    \toprule
    Symbol & Description \\
    \midrule
  $\mathcal{X}$ & point cloud  \\
  $\mathcal{M}$ & metric space\\
  $\mathcal{G}$; V, E & graph; node and edge lists\\
  $\omega(e)$ & weight of edge $e$ \\
  $\mathcal{C}$ & simplicial complex\\
  $\sigma$ & simplex \\
  $\mathscr{C}()$ & clique complex  \\
  $D_a$ & persistence diagram a \\
  $W_r$, $W_{\infty}$ & Wasserstein and Bottleneck  \\
  & distances\\
    L, T  & number of layers and time \\
  & intervals \\
  ${\mathcal{G}^l}$ & graph of a single layer $l$\\
  $\omega^{l^{+}}$ & geodesic edge weight \\
  $\mathbb{D}()$ & distance between \\
  & two persistence diagrams \\
  $\mathcal{G}^{l^{+}}_{t}$ & geodesically densified graph\\
  & of layer $l$ at time $t$ \\
  $d_{GH}$ & Gromov-Hausdorff distance \\
  $\mathbb{F}$ & persistence module \\
  ${d_i()}$ & interleaving distance \\
  $k_{max}$ & maximum clique size  \\
  \bottomrule
  \end{tabular}
     \endgroup
\end{table}

\section{Persistence Methodology for Network Anomaly Detection}
\label{Sec:Method}

We now introduce the new topological method (\texttt{TAD}) for anomaly detection on multilayer graphs and support its design with relevant theoretical guarantees. Table~\ref{tab:notations-tad} details all notations we introduced, and we use the terms graph and network interchangeably.

\begin{definition}[Multilayer network] 
A multilayer network, ${\mathbb{G}} = (\mathcal{G}^{1}, \ldots, \mathcal{G}^{L})$, is a graph structure that consists of $L$ non-overlapping graph layers, where each layer is modeled with a (weighted) graph $\mathcal{G}_i = (V_{i},E_{i},\omega_{i})$, with $i=1,\ldots, L$.
\end{definition}

{\noindent
{\begin{minipage}{1\linewidth}
\textbf{Problem Statement:} 
Let $\{\mathbb{G}_t\}_{t=1}^{T} = \{(\mathcal{G}^{1}_{t}, \ldots, \mathcal{G}^{L}_{t})\}_{t=1}^{T}$ be a $T$ sequence of multilayer networks observed over time $t$, with $1\leq t\leq T<\infty$. The objective is to locate a time point $t^*< T$,
such that an event within the time range $[t^*-m, t^*+m]$, for $0\leq m <t^*$  causes the structure and shape of
$\mathbb{G}_{t^*}$ to differ from the
structural properties of
the earlier observed networks $\mathbb{G}_1, \ldots, \mathbb{G}_{t^*-1}$. With this search, we include anomalies which cause:  \circled{1} the network system to experience a brief shock at $t^*$, and \circled{2} a permanent change in the network system until the next ${t^*} + m$.  
\end{minipage}}
}

\noindent\textbf{Main Idea:} Conceptually, \texttt{TAD} method is designed to associate anomalies in the sequence of multilayer networks to anomalies identified from the time series of their topological summaries. In addition, we introduce our new idea of a specialized persistence diagram for multilayer networks known as the {\em stacked persistence diagram} (SPD). 

\begin{definition}[Stacked Persistence Diagram (SPD)]
\label{SPD}
For a multilayer network ${\mathbb{G} = (\mathcal{G}^{1}, \ldots, \mathcal{G}^{L})}$, we define the associated
PD of $\mathbb{G}$ as
${D}_{\mathbb{G}}$ =(${D}_{\mathcal{G}^1} \bigoplus \ldots \bigoplus$ $ {D}_{\mathcal{G}^L}$), i.e. ${D}_{\mathbb{G}}$ is created as a direct sum of all PDs ${D}_{\mathcal{G}^{l}}$ associated with each single intra-/inter-layer network ${\mathcal{G}^l}\subseteq \mathbb{G}$, for $1\leq l\leq L$.
\end{definition}

\noindent {\bf Why Do We Stack PDs and Why Not to Average PDs?} As our primary focus here is on anomaly detection in multilayer graphs, our goal is to {\it simultaneously capture  joint dynamics} of topological properties exhibited by each graph layer within the interconnected system.  As such, currently existing methods based on averaging PDs and their vectorizations~\cite{munch2015probabilistic,berry2020functional} which are developed for analysis of a single, possibly time-varying object, are not feasible in our context. That is, averaging PDs of the two distinct layers may be viewed as averaging PDs, extracted from apples and oranges. 
In turn, our idea is to jointly track dynamic topological properties which are demonstrated by apple and orange trees over the same time period, and the SPD structure is motivated by the notion of direct sums of multiple vector spaces which serve as mathematical formalization of very different objects.

\begin{figure}[!ht]
  \centering
  \includegraphics[width=0.6\linewidth,height=3.7cm]{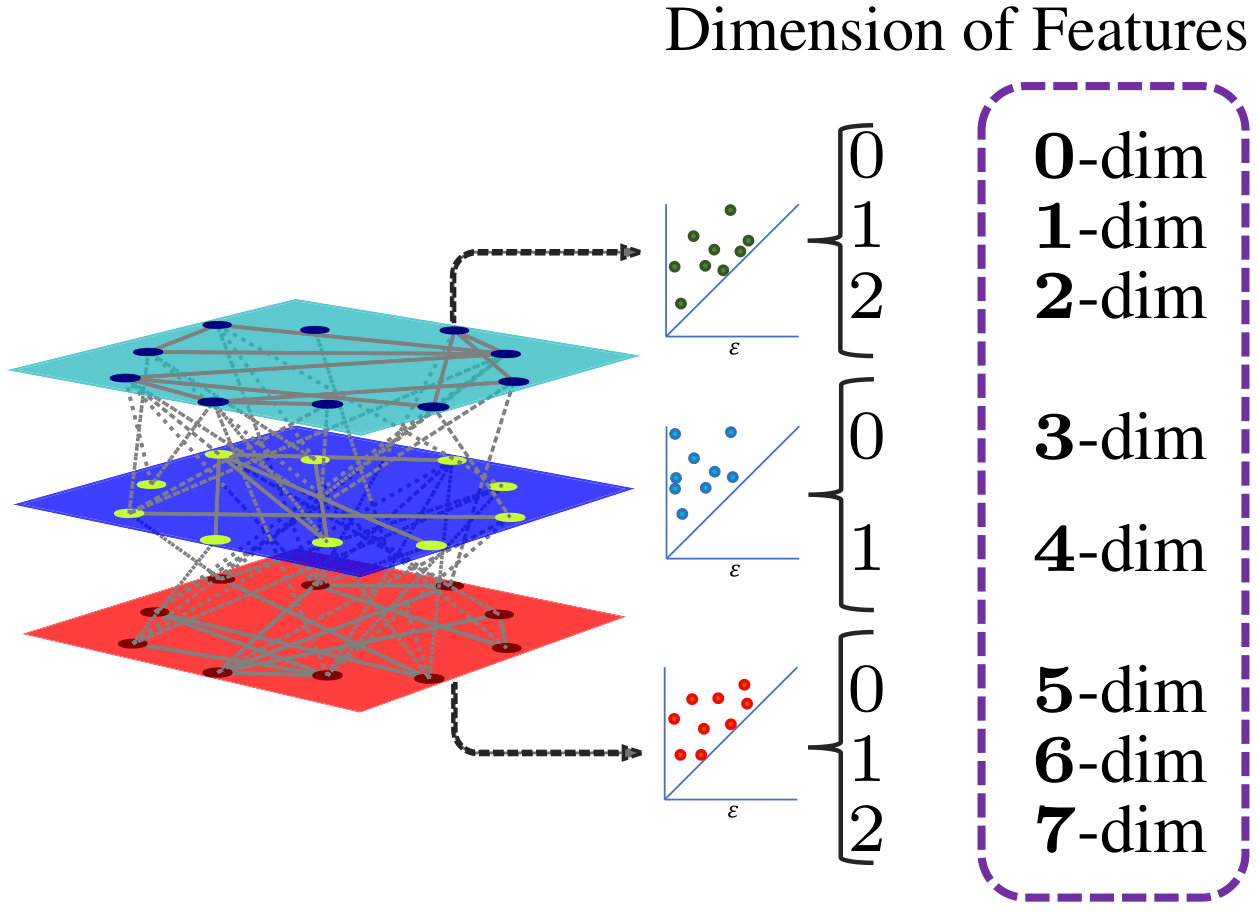}
  \caption{An example of the formation of the SPD for a multilayer network. The multilayer network has 3 layers with PDs that have unequal topological features (3 in the first, 2 in the second, and 3 in the third).  Although the first and third layer PDs contain information about 3-dimensional topological features, they have unequally-positioned points. Essentially, the SPD for the multilayer network will contain information about $7$ classes of topological features.}
  \label{fig:methodMultilayer}
\end{figure}

\noindent{\bf Geodesic Densification of Blockchain Graphs:} Dynamic networks such as Blockchain transaction graphs tend to be sparse, because a node (i.e. an address) can be inexpensively created without proving identity, which allows users to hide their transactions behind new addresses for privacy and security purposes. Furthermore, blockchain communities (e.g. Bitcoin) encourage one-time-use addresses (i.e. creating a new address every time a transaction is created). As a result, a sparse and constantly evolving network structure emerges, making it difficult to rely on conventional network connectivity (i.e. adjacency matrix). To address this limitation, we replace the (weighted) adjacency matrix of the single layer graph $\mathcal{G}^{l}$ of $\mathbb{G}$ with the (weighted) geodesic distance (GD)
matrix~\cite{Biasotti2014AGD} which redefines the edge weights $\omega^{l}$ as $\omega^{l^{+}} = \sum\nolimits_{e\in E(P_{uv})} \omega(e)$, where $P_{uv}$ is the shortest path length between vertex pair $u,v$. This densification reconnects node pairs that have a common path. Paths encode useful information because nodes (i.e. addresses) may merge their coins into a single address to sell them to leave the Blockchain (and thus pay less transaction fees).

The proposed \texttt{TAD} framework operates  according to the following order: 
\[
  \{\mathbb{G}_t\}_{t=1}^{T} \xrightarrow{{T}-step} \{\mathbb{D}(D_{\mathbb{G}_{t-1}},D_{\mathbb{G}_{t}})\}_{t=2}^{T} \xrightarrow{{AD}-step}  
	\{ t_1^*,\ldots\},
\]
where $\mathbb{D}(D_{\mathbb{G}_{t-1}},D_{\mathbb{G}_{t}})$ is any suitable distance metric between two persistence diagrams $D_{\mathbb{G}_{t-1}}$ and $D_{\mathbb{G}_{t}}$. Note that this distance can either be the Bottleneck or the $r$-th Wasserstein distance.

\smallskip
\noindent{\bf T -- step:} At this step, we implement the clique PH to convert the sequence of multilayer networks ${\{\mathbb{G}_t\}_{t=1}^{T}}$ into a sequence of SPDs. This involves the transformation of all the (weighted) adjacency matrices of $\mathcal{G}^{l}_{t}$ into $\mathcal{G}^{l^{+}}_{t}$, followed by the filtration of  persistent topological features by using clique PH.

\smallskip
\noindent{\bf AD -- step:} While \texttt{TAD} method can be integrated with any user-preferred outlier or change point detection algorithm for univariate time series, we 
adopt the recently proposed seasonal extreme studentized deviate test \texttt{S-ESD} ~\cite{Vallis2014AnomalyDetection,Hochenbaum2017AnomalyDetection}. For an observed time series, \texttt{S-ESD} filters out the seasonal component, piecewise approximates the long-term trend component (in order to decrease the instances of false positives) and then incorporates robust statistical learning to identify the location of anomalies. \texttt{S-ESD} is our choice due to its sensitivity to both global anomalies irrespective of seasonal trends and intra-seasonal local anomalies. We provide pseudocode for \texttt{TAD} below, and discuss its computational complexity as well.

\begin{algorithm}[ht]
\label{Alg:TAD}
	\SetKwInOut{Input}{Input}
	\SetKwInOut{Output}{Output}
	\caption{Topological anomaly detection in multilayer networks (\texttt{TAD})}
	\Input{Sequence of $L$-multilayer graphs $\{\mathbb{G}_t\}_{t=1}^{T}  = 
 	\{{{\mathcal{G}_t}^{1},  \ldots, {\mathcal{G}_{t}}^{L}\}^{T}_{t = 1}}$.}
	\Output{Anomalous events $\{{t_1}^{\ast}, \ldots\}$.}
		\For  {t $\leftarrow$ 1 : T}{
			\For  {l $\leftarrow$ 1 : L}{
			Compute GD matrix ${\mathcal{G}_t}^{l^+}$ for ${\mathcal{G}_t}^{l}$ \\
	        Generate the PD  ${\mathcal{D}_{\mathcal{G}_{t}^{l^+}}}$ for ${\mathcal{G}_t}^{l^+}$ \\
			}
			Obtain SPD $\mathcal{D}_{\mathbb{G}_{t}}$ by chronologically stacking PDs from  ${\mathcal{D}_{\mathcal{G}_{t}^{1^+}}}$ to ${\mathcal{D}_{\mathcal{G}_{t}^{L^+}}}$
		}
		
	\For {t $\leftarrow$ 2: T}{
	With suitable distance metric ($\mathbb{D}$), obtain similarity between $\mathcal{D}_{\mathbb{G}_{t-1}}$ and $\mathcal{D}_{\mathbb{G}_{t}}$
	}
	With \texttt{S-ESD}, detect anomalies (${t_1}^{\ast},\ldots$) from the series \big\{${\mathbb{D}}(\mathcal{D}_{\mathbb{G}_{1}},\mathcal{D}_{\mathbb{G}_{2}}),\ldots,$\\ ${\mathbb{D}}(\mathcal{D}_{\mathbb{G}_{T-1}},\mathcal{D}_{\mathbb{G}_{T}})$\big\} 
	
\end{algorithm}

\subsection*{Computational Complexity for $\texttt{TAD}$ procedure} 
Algorithm~1 details the computation of PDs based on finding and merging clique communities. The available clique implementation of~\cite{Rieck2017clique} (which we utilize here) is highly efficient and has a complexity of $\mathcal{O}(\kappa \varphi^{-1}(\kappa) )$ for persistent homology computation, where $\kappa$ is the number of edges and $\varphi^{-1}(\cdot)$ is the extremely slow-growing inverse of the Ackermann function~\cite{Ackermann:Grossman:1988}. It works jointly with one of the many clique percolation algorithms~\cite{Fortunato201075}. To compute the Wasserstein distance metric, we use the efficient polynomial time algorithm~\cite{NoceWrig06}.

\subsection{Theoretical Properties of the Stacked Persistence Diagram}
As shown by~\cite{chazal2008stability}, the conventional PD of an object (i.e. a single layer graph or point cloud) is stable under minor data perturbations. Noting that SPD is derived from the direct sum of the persistence modules corresponding to each layer in $\mathbb{G}$ and using the Isometry theorem for individual persistence modules~\cite{chazal2016structure}, we derive similar theoretical guarantees for SPD.

    \begin{theorem}[Stability of  SPD]
    \label{theorem}
    Let $\mathbb{G}_{X} = \{{\mathcal{G}_{X}^1},\ldots,{\mathcal{G}_{X}^L}\}$ and $\mathbb{G}_{Y} = \{{\mathcal{G}_{Y}^1},\ldots,$ ${\mathcal{G}_{Y}^L}\}$ be two multilayer networks generated from the same space of $L$-multilayer networks. Then
    \begin{displaymath}
        W_{\infty} ({D_{\mathbb{G}_{X}}},{D_{\mathbb{G}_{Y}}} ) \leq \max_{1\leq l\leq L} \bigl( d_{GH}\bigl( \{{\mathcal{G}_{X}^l}, {\omega_{\mathcal{G}_{X}^l}}\},\{{\mathcal{G}_{Y}^l}, {\omega_{\mathcal{G}_{Y}^l}}\}\bigr)\bigr)
    \end{displaymath}
 \noindent 
 where ${W_{\infty}}$  is the Bottleneck distance and ${d_{GH}}$ is the Gromov-Hausdorff distance.
\end{theorem}

\noindent Proof for Theorem~\ref{theorem} is in the Appendix. Theorem~\ref{theorem} implies that the proposed new SPD $D_{\mathbb{G}}$ (see Definition~\ref{SPD}) for any multilayer network $\mathbb{G}$ is robust with respect to $W_{\infty}$ under minor input data perturbations. As a result, 
Theorem~\ref{theorem} provides theoretical foundations to our \texttt{TAD} idea. Hence,
under the null hypothesis of no anomaly, we expect to observe similar SPDs over dynamic multilayer networks $D_{\mathbb{G}_t}$, while a noticeable difference between two adjacent SPDs is likely to be a sign of anomaly. Note that stability of SPD in terms of $W_{1}$ requires vectorization of SPD and Lipschitz continuity of the associated vectorization. While such vectorization approaches are highly successful for image and graph learning (see, e.g.,~\cite{adams2017persistence,hofer2019learning,zhao2019learning}), our preliminary studies show lack of sensitivity of such vectorization techniques in conjunction with network anomaly detection.  

\section{Experiments on Blockchain Networks}
\label{Sec:Experiments}

\subsection{Experimental Setup}

{\bf Baseline Algorithms:} We compare performance of \texttt{TAD} method against the following strong state-of-the-art (SOTA)  algorithms for anomaly detection on single layer networks: \circled{1} DeltaCon by \cite{Koutra2016deltacon} (which we label DC) for weighted/unweighted networks, \circled{2} Scan Statistics algorithm by \cite{Chen2015scanstat} (which we label gSeg) for unweighted networks, \circled{3} Edge monitoring method with Euclidean distance by \cite{Wang2017fast} (which we label EMEu) for weighted networks, and \circled{4} Edge monitoring method with Kullback-Leibler divergence by \cite{Wang2017fast} (which we label EMKL) for weighted networks. 
Finally, we also considered an embedding-based algorithm for anomaly detection. That is, we tracked Frobenius norms among embeddings of multilayer blockchain graphs at each time snapshot, delivered by the one of the most widely used algorithms for multilayer graph embedding, MANE of~\cite{li2018multi}.
This \circled{5}-th approach is denoted by Graph-Em. We provide a brief description of the mechanism for each method in the Appendix. For all competing methods, we use the default parameters reported in the literature. Wherever applicable, we set a standard level of significance $\alpha$ of 0.05.

Since all competing methods are designed for single layer networks, we implement them (individually) w.r.t. each $l$ layer in all the multilayer graphs $\{\mathbb{G}_{t}\}_{t=1}^{T}$ and then combine the detected results, while correcting for the multiple hypothesis testing framework. In the Appendix, we provide two types of multiple hypotheses that specifies how we retain anomalies for the sequence of multilayer graphs and these include: \circled{1} keep all anomalies identified from at least one $\{\mathcal{G}^{l}_{t}\}_{t=1}^{T}$, \circled{2} keep all anomalies that are commonly identified from all $\{\mathcal{G}^{l}_{t}\}_{t=1}^{T}$. We provide results for choice (1), and defer the results for (2) to the Appendix. Additionally, we construct a single layer version of $\texttt{TAD}$ (which we call $\texttt{S-TAD}$) and apply this to the same single layers. To be precise, our improvised \texttt{S-TAD} will extract PDs from each $l$-layer, and without creating SPDs, apply the chosen distance metric to consecutive PDs to obtain a time series of topological summaries for the sequence $\{\mathbb{G}^{l}_{t}\}_{t=1}^{T}$.  Therefore, our evaluation will investigate the performance of \texttt{TAD} method against the performance of the chosen techniques (DC, gSeg EMEu, EMKL, Graph-Em), and  $\texttt{S-TAD}$  when the (un)weighted multilayer networks are viewed as a multiple hypothesis. 

\begin{figure*}
\centering
\includegraphics[height = 0.3\textheight,keepaspectratio]{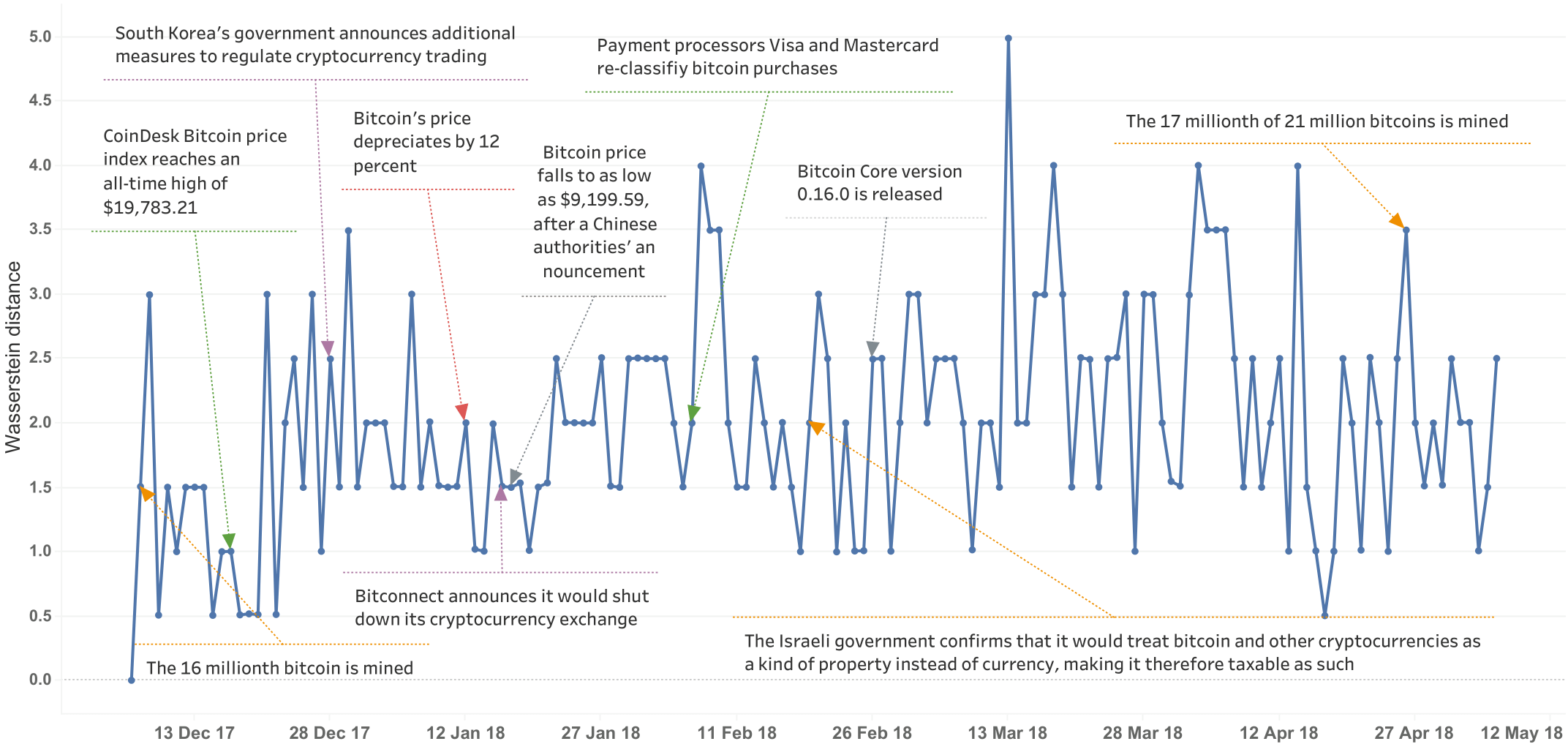}
\caption{Anomalous events detected by \texttt{TAD} for the multilayer Ethereum network.}
\label{fig:EventMultilayerEthereum}
\end{figure*}

\noindent{\bf Topological Distances in \texttt{TAD}} We have experimented with various topological metrics, particularly, $W_{\infty}$ bottleneck  and $W_1$ Wasserstein  distances. While our preliminary results do not indicate that $W_1$ yields substantial gains over $W_{\infty}$ (i.e. 70\% of the true anomalous events are detected regardless of the distance choice), $W_1$ tends to be slightly more sensitive than $W_{\infty}$. As such, we proceed with $W_1$ as the primary choice and consider $W_{1}(D_{\mathbb{G}_{t-1}}, D_{\mathbb{G}_{t}})$, between consecutive SPDs $D_{\mathbb{G}_{t-1}}$ and $D_{\mathbb{G}_{t}}$ for $2\le t \le T$.  We apply the \texttt{TAD} technique to two input data types: weighted and unweighted multilayer networks. Edge weight is defined as a number of transactions between nodes. 

\noindent{\bf Reproducibility and replicability} The anonymized codes and data sets for this project are available at \url{https://github.com/tdagraphs}.

\subsection{Ethereum Token Networks}
\label{Sec:ExperimentsTokens}
{\bf Data set:}  The Ethereum blockchain was created in 2015 to implement Smart Contracts, which are Turing complete software codes that execute user defined tasks. Among many possible tasks, contracts are used to create and sell digital assets on the blockchain. The assets can be categorized into two categories: \circled{1} Tokens whose prices can fluctuate; ERC20 or ERC721~\cite{victor2019measuring}, \circled{2} Stablecoins whose prices are pegged to an asset such as USD~\cite{moin2019classification} (these are also ERC20 tokens). Token networks are particularly valuable because each token naturally represents a network layer with the same nodes (addresses of investors) appearing in the networks (layers) of multiple tokens. For our experiments, we extract token networks from the publicly available Ethereum blockchain, and use the normalized number of transactions between nodes as the edge weights. By principle, a token network is a directed, weighted multigraph where an edge denotes the transferred token value. Although address creation is cheap and easy, most blockchain users use the same address over a long period. Furthermore, the same address may trade multiple tokens. As a result, the address appears in networks of all the tokens it has traded. From our data set timeline, we only include tokens reported by the \url{EtherScan.io} online explorer to have more than \$100M in market value. Eventually, the data set contains 6 tokens, and on average, each token has a history of 297 days (minimum and maximum of 151 and 576 days, respectively). Note that each token has a different creation date, hence token networks  have non-identical lifetime intervals.

\begin{table}[t]
  \caption{Anomaly detection performance for the weighted Ethereum blockchain and Ripple currency networks.}
  \label{table_EtherRip}
  \centering
  \begingroup
  \setlength{\tabcolsep}{2.8pt} 
  \begin{tabular}{llllllllllll}
    \hline
     &  \multicolumn{5}{c}{Ethereum} &  & \multicolumn{5}{c}{Ripple} \\
    \cmidrule(r){2-6} \cmidrule(r){8-12}
     & \texttt{S-TAD}  & DC & EMEu &  EMKL & \texttt{TAD} &  & \texttt{S-TAD}  & DC & EMEu &  EMKL & \texttt{TAD}\\
    \hline
     TP   & 15    & 52   & 3   & 5  & 10 &  &  95  & 105    & 10    &  10   & 16 \\
     FP   & 28    & 69   & 3   & 5 & 2   &  &  260  & 283    & 40    &  32   & 9 \\
     TN   & 99    & 30   & 132  & 130 & 135  &  &  872  & 837    & 1152  & 1161  & 1187 \\
     FN   & 10    & 1    & 14   & 12 & 5 &  &  35  & 37     & 60   & 59   & 50 \\
     Acc. & 0.750  & 0.539 & 0.888 & 0.888 & {\bf 0.954} &  & 0.766 & 0.746  & 0.921 & 0.928 & {\bf 0.953} \\
   \hline
  \end{tabular}
  \endgroup
\end{table}

\begin{figure*}[b]
\vspace{-5mm}
        \centering
        \begin{subfigure}{0.49\textwidth}
            \centering  
        \includegraphics[width=1.1\columnwidth, height = 4.1cm]{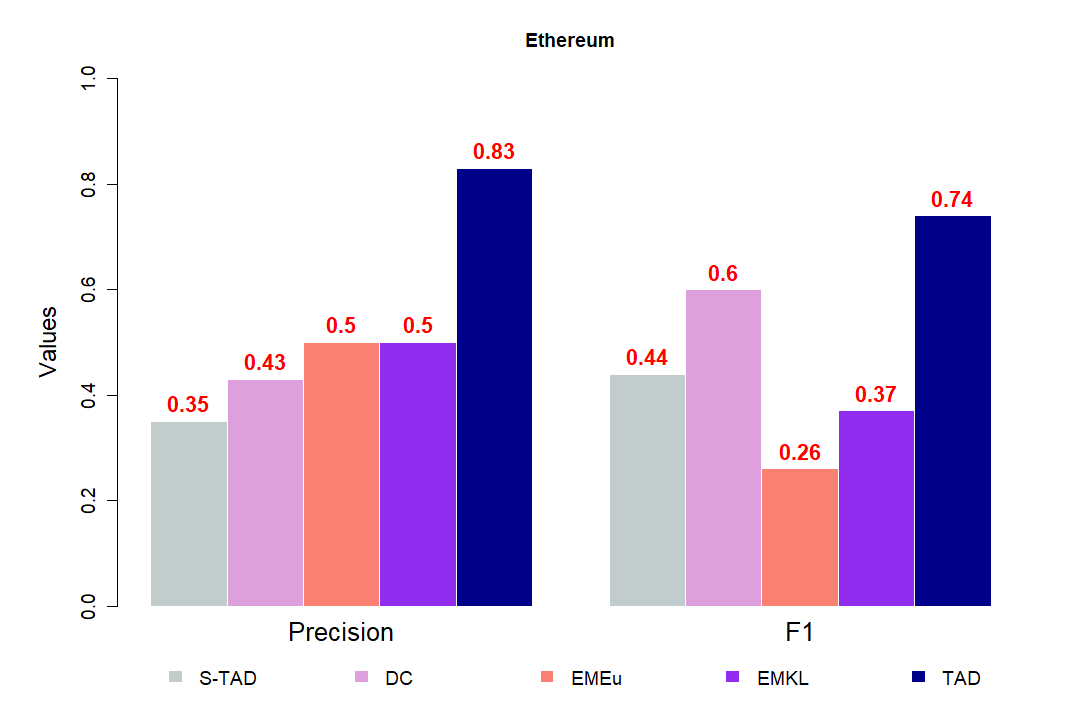}
        \end{subfigure}
            \begin{subfigure}{0.49\textwidth}  
            \centering 
             \includegraphics[width=1.1\columnwidth, height = 4.1cm]{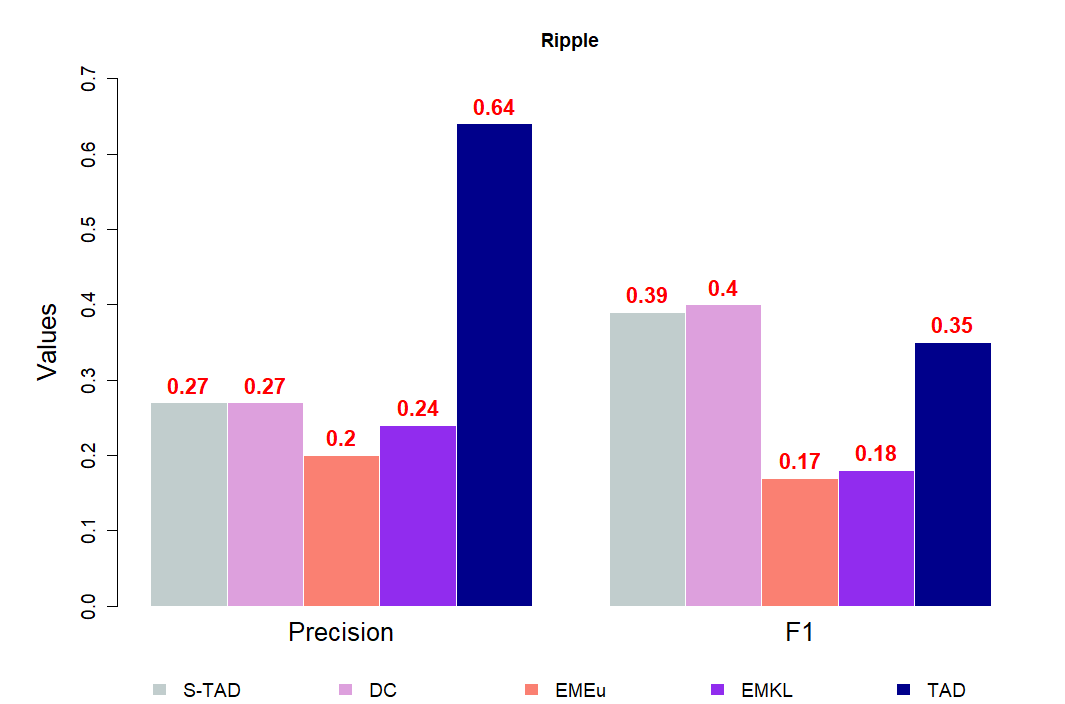}
         \end{subfigure}
        \caption{Precision and F1 scores for the weighted Ethereum blockchain and Ripple currency networks.}
        \label{fig:f1EtherRip}
    \end{figure*}

\noindent{\bf Ground Truth:} As ground truth, we adopt and curate Blockchain events from Wikipedia~\cite{Bitcoin_global_events}, which lists and explains major events since 2008. In total, there are 72 events that have shaped blockchain networks --- some of them in adverse (see the supplementary material for the complete list). However, token networks cannot detect events before 2015 because Ethereum and its tokens did exist before then. Hence, our experiments focused on at most 32 (out of the 72) the token transaction events.

\noindent{\bf Results:} Table~\ref{table_EtherRip} presents summary statistics for the weighted token multilayer network analysis against the three single-layer SOTA solutions (i.e. DC, EMEu, EMKL) and our topological \texttt{S-TAD} method. We find that \texttt{TAD} delivers lower FP values. In addition, we notice that \texttt{TAD} achieves a significantly higher accuracy ($> 7\%$ of what EMEu/EMKL report). This is evidenced by the detected points in Figure~\ref{fig:EventMultilayerEthereum}. From Figure~\ref{fig:f1EtherRip} we notice again that the \texttt{TAD} yields substantially higher Precision ($66\%$ more than what DC gets) and F1 ($> 23\%$ of what DC gets) values, implying that \texttt{TAD} tends to be substantially more efficient in locating  relevant anomalies within the multilayer graph sequence than its competitors. In addition, we find that the performance results of the Graph-Em method in Table~\ref{table_Ether_unw} and Figure~\ref{fig:f1EtherRip_uw} are substantially worse than the ones delivered by our proposed \texttt{TAD}. This phenomenon can be explained by higher data aggregation typically performed by graph embedding tools which results in lower sensitivity to anomalous changes in the graph structure. Altogether, these results indicate that \texttt{TAD} tends to be the most preferred tool for identifying anomalies in the multilayer network setting. 
Table~\ref{table_Ether_unw} presents experimental results for the anomalous event detection in the unweighted multilayer Ethereum blockchain networks. We find that \texttt{TAD} delivers the highest detection accuracy (0.954, which is about 20\% greater than what gSeg yields). In addition, we notice that \texttt{TAD} attains the lowest FP value (about 10\% of what gSeg obtains) and the highest TN value (about 27\% more than what gSeg gets). In turn, Figure~\ref{fig:f1EtherRip_uw} suggests that \texttt{TAD} yields the highest precision (93\% greater than DC) and the highest F1 score (23\% more than DC). These findings suggest that the new \texttt{TAD} method tends to be the most accurate approach for flagging relevant anomalous events.

\begin{table}[t]
  \caption{Anomaly detection performance for the unweighted Ethereum blockchain networks.}
  \label{table_Ether_unw}
  \centering
  \begingroup
  \setlength{\tabcolsep}{2.8pt} 
  \begin{tabular}{llllllllllll}
    \hline
      &  \multicolumn{5}{c}{Ethereum} &  & \multicolumn{5}{c}{Ripple}                   \\
    \cmidrule(r){2-6} \cmidrule(r){7-12}
    & \texttt{S-TAD}  & DC & gSeg &  Graph-Em & \texttt{TAD} & & \texttt{S-TAD}  & DC & gSeg &  Graph-Em & \texttt{TAD}\\
      \hline
     TP   & 17    & 52     & 14    & 3     & 10          & & 80    & 105    & 17    & 0 & 11 \\
     FP   & 28    & 69     & 21    & 11    & 2           & & 241   & 283    & 56    & 1 & 23\\
     TN   & 97    & 30     & 106   & 126   & 135         & & 900   & 837    & 1130  & 1195 & 1173\\
     FN   & 10    & 1      & 11    & 12    & 5           & & 41    & 37     & 59    & 66 & 55\\
     Acc. & 0.750  & 0.539 & 0.789 & 0.849 & {\bf 0.954} & & 0.777 & 0.746  & 0.909 & {\bf 0.947}  &  0.938 \\
       \hline
  \end{tabular}
  \endgroup
\end{table}

\subsection{Ripple Currency Networks}
\label{Sec:ExperimentsRipple}
{\bf Data set:} The Ripple Credit Network was created to facilitate remittance across countries, but the network has transitioned to a blockchain-like structure where network approved entities (e.g., banks) issue currencies in I-Owe-You notes, and addresses can trade these currencies in blocks (which are called ledgers). On the Ripple network any real life asset, such as Chinese Renminbi or  US Dollar, can be issued by certain participants only but traded by all addresses (nodes). In terms of regulatory issues by governments and price movements, Ripple is a part of the Blockchain ecology and the networks are impacted by the global events such as government regulations and trade volume increases~\cite{moreno2018mind}. We use the official Data API (\url{https://xrpl.org/data-api.html}) and extract the five most issued fiat currencies on the Ripple network: JPY, USD, EUR, CCK, CNY. We construct a multilayer network from the payment transactions of the five currencies that covers a timeline of Oct-2016 to Mar-2020. Similar to the Ethereum token analysis, we use the normalized number of transactions between nodes as the edge weights.

\noindent{\bf Ground Truth:} As ground-truth, we use the same events described in the Ethereum token network experiments.
However, since the Ripple data set has a longer temporal span of observations than the Ethereum token networks, there are a total of 66 Blockchain events.

\begin{figure*}[t]
\vspace{-5mm}
        \centering
        \begin{subfigure}{0.49\textwidth}
            \centering  
        \includegraphics[width=1.1\columnwidth, height = 4.1cm]{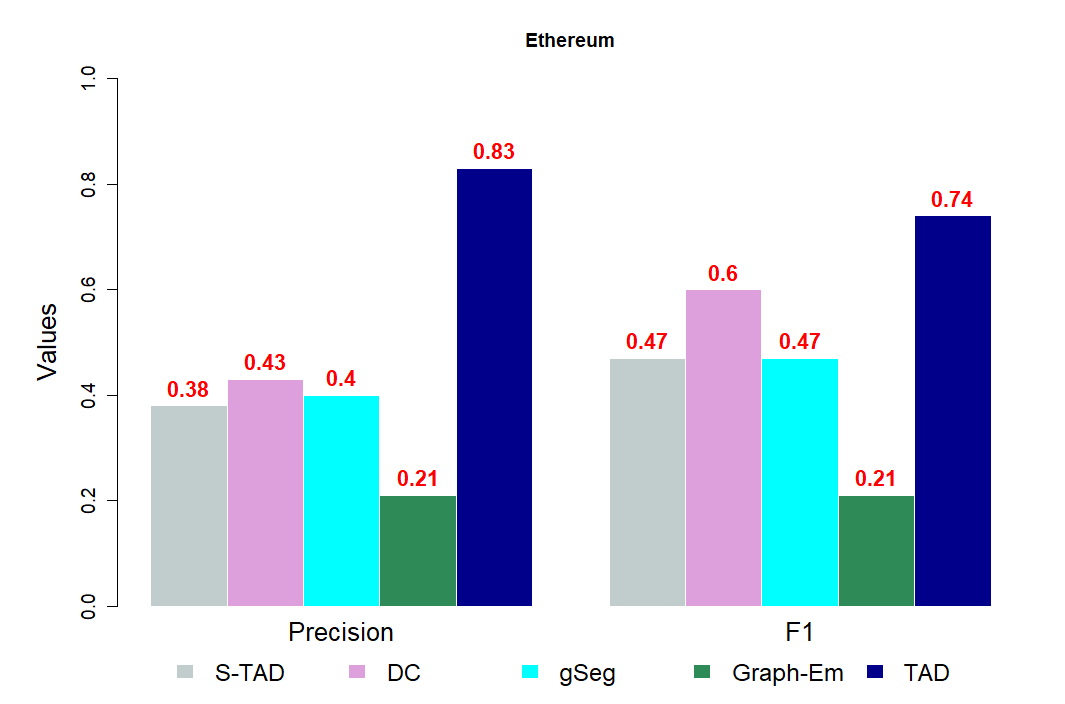}
            \end{subfigure}
        \begin{subfigure}{0.49\textwidth}  
            \centering 
            \includegraphics[width=1.1\columnwidth, height = 4.1cm]{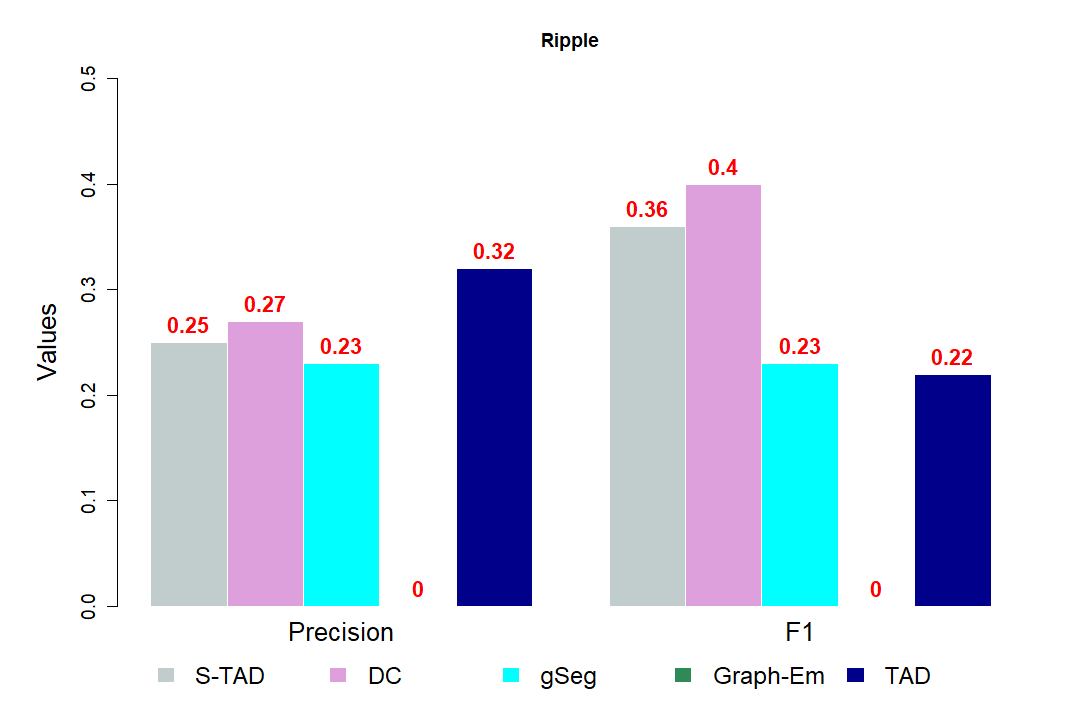}
        \end{subfigure}
        \caption{Precision and F1 scores for the unweighted Ethereum blockchain and Ripple currency networks.}
        \label{fig:f1EtherRip_uw}
    \end{figure*}

\noindent{\bf Results:} Summaries from Table~\ref{table_EtherRip} indicate that \texttt{TAD} attains the highest event detection accuracy (0.953). Furthermore, we find that \texttt{TAD} yields the lowest FP value, which is actually 22.5\% of the value by EMEu and about 28\% of the value by EMKL.  Figure~\ref{fig:f1EtherRip} displays detection results for the anomalous events in the multilayer Ripple payment networks. We find that \texttt{TAD} yields the highest precision (more than double what DC/\texttt{S-TAD} get) and is close to the top F1 score. Differing from Ethereum experiment, 
the best F1 performance is delivered by DC, closely followed by \texttt{S-TAD} and then \texttt{TAD}. 
Table~\ref{table_Ether_unw} suggests that \texttt{TAD} delivers the highest detection accuracy (0.938, which is about 3\% greater than what gSeg yields) for the unweighted Ripple currency network. In addition, we notice that \texttt{TAD} attains the lowest FP value (about 41\% of what gSeg obtains) and the highest TN value (about 3\% more than what gSeg gets). In turn, Figure~\ref{fig:f1EtherRip_uw} shows that \texttt{TAD} yields the highest precision (about 18\% greater than DC) but the lowest F1 score (55\% of what DC gets).

Finally, note that in Ethereum we use 6 tokens, whereas Ripple experiments are performed on 5 currencies. As Tables~\ref{table_Ether_unw}
and~\ref{table_EtherRip} suggest, the Ethereum results appear to be better than those of Ripple. That is, detection accuracy substantially improves with a higher number of layers. However, for both cases  \texttt{TAD} either outperforms or on par with baseline techniques. The key intuition behind these results is that \texttt{TAD} allows for simultaneous evaluation of subtle changes in multiple homological features both within network layers and across network layers in sparse dynamic environments of blockchain transaction graphs. As such, SPD appears to be more sensitive to subtle changes in the multilayer network structure than competing non-TDA tools.

\section{Conclusion}
\label{Sec:Conclus}We have proposed the first topological anomaly detection (\texttt{TAD}) framework for dynamic multilayer networks. We have derived stability guarantees
of the new topological summary for multilayer graphs, i.e., stacked persistence diagram, which is the key tool behind \texttt{TAD}  and validated utility of \texttt{TAD} on two blockchain transaction graphs. Our studies have indicated that \texttt{TAD} yields a highly competitive performance in detecting anomalous events on Ethereum and Ripple blockchains. In the future we plan to advance \texttt{TAD} to anomaly detection in attributed dynamic networks and analysis of evolving communities.

\section*{Appendix}
\label{Sec:Appendix}

\subsection*{Theoretical proof of Theorem~1}
To prove Theorem 4.3, we invoke the following arguments. First, we note a relationship between conventional PD of a single layer network and its associated persistence module.
Then the results on isometry of PDs and 
persistent modules allow us to shift our focus from stability of PDs to stability of the associated persistent modules.
Second, as no two layers in
the multilayer network $\mathcal{G}$ share edges, the persistence module associated with SPD is then a direct sum of persistence modules associated with PDs of single layer networks.
Finally, armed with the results of~\cite{chazal2009gromov,chazal2016structure} on stability of persistence modules and accounting for the direct sum representation of SPD, we arrive to stability of SPD. 

Now we state the key mathematical definitions and results which are required for our derivations.

\vspace{0.5em}
\setcounter{section}{1}

\begin{theorem}\textbf{(Stability of PD,~Theorem 3.1 in \cite{chazal2009gromov})}
For any finite metric spaces \{X,${w_X}$\} and \{Y,${w_Y}$\},
\begin{displaymath}
        W_{\infty} ({D_X},{D_Y}) \leq d_{GH}\Big( \{{X},{w_{X}}\},\{{Y},{w_{Y}}\}\Big), 
\end{displaymath}
where $d_{GH}$ is the Gromov-Hausdorff distance~\cite{chazal2009gromov}.
\end{theorem}

\noindent Note that weight functions $\omega_{X}$ and $\omega_{Y}$ correspond to pairwise node distances within networks $X$ and $Y$ respectively.

\begin{definition}[Persistence module]
A persistence module $\mathbb{F}=\{{F_a},{\mathrm{}{f}_a}^b\}$ is an indexed family of vector spaces 
(${F_a} \mid a \in \mathbb{R}$) together with the double-indexed family of linear maps (${\mathrm{}{f}_a}^b:F_{a}\rightarrow F_{b} \mid a\leq b$) which satisfy the composition law 
${\mathrm{}{f}_a}^c= {\mathrm{}{f}_a}^b \circ  {\mathrm{}{f}_b}^c$  whenever $a\leq b\leq c$ and where ${\mathrm{}{f}_a}^a$ is  the identity map of $F_a$~\cite{chazal2016structure}.
\end{definition}

\noindent A persistence module $\mathbb{F}=\{{F_a},{\mathrm{}{f}_a}^b\}$ is $\textbf{q-tame}$ if rank (${\mathrm{}{f}_a}^b)<\infty$  for any $a< a + q <b$~\cite{chazal2008stability}. 

\begin{definition}[Interleaving distance]
For $\delta\geq 0$, the two modules $\mathbb{F}=\{{F_a},{\mathrm{}{f}_a}^b$\}  and $\mathbb{M}=\{{M_a},{\mathrm{}{m}_a}^b$\}  are \textbf{$\delta$-interleaved} if for all $a\in \mathbb{R}$  there exists the collection of linear maps
${\Phi}: {F_a} \rightarrow {M_{a+\delta}}$ and ${\Theta: M_a  \rightarrow F_{a+\delta}}$ such that all diagrams that can be composed out of the maps ${\Phi_a}, {\Theta_a}$, and the linear maps of\:  $\mathbb{F}$ and $\mathbb{M}$ commute~\cite{chazal2009proximity,bubenik2015statistical,bjerkevik2019computing}. 

\noindent Hence, the interleaving distance ${d_i}$ between $\mathbb{F}$ and $\mathbb{M}$ is defined as 
\[{d_i(\mathbb{F}, \mathbb{M})} = \inf({\delta} \mid \mathbb{F} \text{ and } \mathbb{M}\text{ are }{\delta}-interleaved),\]  
where $\inf$ represents the infimum.
\end{definition}

\vspace{1em}
\begin{theorem}\textbf{(Isometry Theorem,~Theorem 4.11 in \cite{chazal2016structure})}
Let $\mathbb{F}$ and $\mathbb{G}$ be {\em $q$-tame} persistence modules. Then
\begin{displaymath}
         d_{i}(\mathbb{F}, \mathbb{G}) = W_{\infty}({D}_{F}, {D}_{G}).
\end{displaymath}
\end{theorem}

\vspace{1em}
\begin{prop}[Proposition 4.5 in \cite{chazal2016structure}]
Let $\mathbb{U}_{1}$,  $\mathbb{U}_{2}$,  $\mathbb{V}_{1}$ and  $\mathbb{V}_{2}$ be persistence modules. Then
{
\begin{displaymath}
        {d_i}({\mathbb{U}_{1} \oplus \mathbb{U}_{2}},{\mathbb{V}_{1} \oplus \mathbb{V}_{2}}) \leq \max \Big({d_i}(\mathbb{U}_{1},\mathbb{U}_{2}),{d_i}(\mathbb{V}_{1}, \mathbb{V}_{2}) \Big).
\end{displaymath} } 
Generally, let ($\mathbb{U}_{s} \mid s \in S$) and ($\mathbb{V}_{s} \mid s \in S$) be families of persistence modules indexed by the same set S, and let
\begin{displaymath}
         \mathbb{U} = \bigoplus\limits_{s \in S}\mathbb{U}_{s}\:\:\:\: \text{and} \:\:\:\:\mathbb{V} = \bigoplus\limits_{s\in S}\mathbb{V}_{s}. 
\end{displaymath}
Then
{
\begin{displaymath}
        {d_i}({\mathbb{U}, \mathbb{V}}) \leq \sup_{s \in S} \Big({d_i}(\mathbb{U}_{s},\mathbb{V}_{s})\mid s \in S \Big),
\end{displaymath} }
where $\oplus$ 
denotes a direct sum.

\end{prop}

{\bf Proof of Theorem~1}
Let $\mathcal{G}_{X} = (V_{X}, \omega_{X})$ and $\mathcal{G}_{Y} = (V_{Y}, \omega_{Y})$ be two single layer networks equipped with clique filtrations $\mathscr{F}_X$ and $\mathscr{F}_Y$ respectively. 

Armed with $\mathscr{F}_X$ and $\mathscr{F}_Y$, we generate the family of persistence modules $\mathbb{F}_X$ and $\mathbb{F}_Y$, respectively~\cite{chazal2008stability,chazal2016structure}. As discussed by~\cite{bubenik2018topological,de2013geometry}, $\mathbb{F}_X$ and $\mathbb{F}_Y$ induce the corresponding PDs $D_{\mathcal{G}_X}$ and $D_{\mathcal{G}_Y}$. Hence, the topological persistence patterns for $\mathcal{G}_{X}$ and $\mathcal{G}_{Y}$ are
\[\mathcal{G}_{X}\rightarrow \mathscr{F}_X\rightarrow {\mathbb{F}_X}\rightarrow{D_{\mathcal{G}_X}}
\:\:\:\:\:\:\:\:\:\:\:\:\:\:\:\text{and}\:\:\:\:\:\:\:\:\:\:\:\:\:\:\:
\mathcal{G}_{Y}\rightarrow \mathscr{F}_Y \rightarrow{\mathbb{F}_Y}\rightarrow{D_{\mathcal{G}_Y}}.\]
By assumptions, $\mathbb{F}_X$ and $\mathbb{F}_Y$ are $q$-tame. 
Hence, in view of the Isometry theorem (Theorem~4.11 of \cite{chazal2016structure}), we can invoke interleaving distance $d_{i}$
to assess relationship between $d_{i}$ and $W_{\infty}$. 
Under a Vietoris-Rips filtration (of which clique filtration is a subcase), the Stability of PDs theorem  (Theorem~3.1 of~\cite{chazal2009gromov}) implies that 
\begin{equation}
\label{eq1}
        W_{\infty} ({D_{\mathcal{G}_X}},{D_{\mathcal{G}_Y}}) \leq d_{GH}\Big( \{{\mathcal{G}_X},{\omega_{\mathcal{G}_X}}\},\{{\mathcal{G}_Y},{w_{\mathcal{G}_Y}}\}\Big), 
\end{equation}
where $d_{GH}$ is the Gromov-Hausdorff distance. By $({\mathcal{G}_X},{\omega_{\mathcal{G}_X}})$ we mean the induced finite metric space $(V_{X}, \omega_{X})$ where the distances between the finite vertex set is induced by the edge weight function $\omega_{X}$. Similar reference is implied for $({\mathcal{G}_Y},{\omega_{\mathcal{G}_Y}})$. In turn, the Isometry theorem implies
\begin{displaymath}
        {d_i}({\mathbb{F}_X},{\mathbb{F}_Y}) = W_{\infty}({D_{\mathcal{G}_X}},{D_{\mathcal{G}_Y}}).
\end{displaymath}           Hence, in view of~\eqref{eq1}, we obtain
\begin{equation}
\label{eq2}
        {d_i}({\mathbb{F}_X},{\mathbb{F}_Y}) \leq d_{GH}\Big( \{{\mathcal{G}_X},{\omega_{\mathcal{G}_X}}\},\{{\mathcal{G}_Y},{w_{\mathcal{G}_Y}}\}\Big).
\end{equation}

\vspace{1em}
Now consider two multilayer networks $\mathbb{G}_{X} = \{\mathcal{G}_{X}^{1},\ldots,\mathcal{G}_{X}^{L}\}$ and $\mathbb{G}_{Y} = \{\mathcal{G}_{Y}^{1},\ldots,\mathcal{G}_{Y}^{L}\}$, with $\omega_{\mathbb{G}_{X}}= \{\omega_{\mathcal{G}_{X}^1}, \ldots, \omega_{\mathcal{G}_{X}^{L}}\}$ and $\omega_{\mathbb{G}_{Y}}= \{\omega_{\mathcal{G}_{Y}^1}, \ldots, \omega_{\mathcal{G}_{Y}^{L}}\}$ as the weight functions. Clearly, $\{\mathbb{G}_{X}, \omega_{\mathbb{G}_{X}}\}$ and $\{\mathbb{G}_{Y}, \omega_{\mathbb{G}_{Y}}\}$ are metric spaces. From the definition of SPD (Definition 3 under ``Persistence Methodology for Network Anomaly Detection''), $\mathbb{G}_X$ and $\mathbb{G}_Y$ induce SPDs  \[D_{\mathbb{G}_X}=(D_{\mathcal{G}_{X}^{1}},\ldots,D_{\mathcal{G}_{X}^{L}})\:\:\:\:\:\:\:\:\:\:\:\:\:\:\:\text{and}\:\:\:\:\:\:\:\:\:\:\:\:\:\:\: D_{\mathbb{G}_Y}=(D_{\mathcal{G}_{Y}^{1}}\ldots,D_{\mathcal{G}_{Y}^{L}})\] respectively, and based on \eqref{eq1}  and \eqref{eq2} above, for $l\in \{1,\ldots,L\}$
\begin{displaymath}
        W_{\infty} ({D_{\mathcal{G}_{X}^{l}}},{D_{\mathcal{G}_{Y}^{l}}}) \leq d_{GH}\Big( \{{\mathcal{G}_{X}^{l}},{\omega_{\mathcal{G}_{X}^{l}}}\},\{{\mathcal{G}_{Y}^{l}},{\omega_{\mathcal{G}_{Y}^{l}}}\}\Big),
\end{displaymath}
    which implies that                             
\begin{equation}
\label{eq3}
        {d_i}({\mathbb{F}_{X^l}},{\mathbb{F}_{Y^l}}) \leq d_{GH}\Big( \{{\mathcal{G}_X^l},{\omega_{\mathcal{G}_X^l}}\},\{{\mathcal{G}_Y^l},{\omega_{\mathcal{G}_Y^l}}\}\Big).
\end{equation}  

By SPD construction, 
no edges are shared between any 
$\mathcal{G}_X^i$ and $\mathcal{G}_X^j$, $i\ne j$. Hence, the persistence module of $\mathbb{G}_X$
corresponding to $D_{\mathbb{X}}$ is the direct sum of all $\mathbb{F}_{X^l}$:
\begin{equation}
\label{direct_sum}
        \mathbb{F}_{\mathbb{X}} = \bigoplus\limits_{1\leq l \leq L} \mathbb{F}_{X^l}.
\end{equation}

According to Proposition~1 above (i.e. Proposition~4.5 in \cite{chazal2016structure}),  
\begin{equation}
\label{eq4}
        {d_i}({\mathbb{F}_{\mathbb{X}}},{\mathbb{F}_{\mathbb{Y}}}) \leq \max_{1 \leq l \leq L}\Big({d_i}({\mathbb{F}_{X^l}},{\mathbb{F}_{Y^l}})\Big),
\end{equation}  
and by assumptions, each $\mathbb{F}_{X}^{i}$ and $\mathbb{F}_{Y}^{j}$ is $q$-tame, $i\ne j$. Combining~\eqref{eq3} and~\eqref{eq4} results in
\begin{displaymath}
         {d_i}({\mathbb{F}_{\mathbb{X}}},{\mathbb{F}_{\mathbb{Y}}}) \leq \max_{1 \leq l \leq L}\Big(d_{GH}\Big( \{{\mathcal{G}_X^l},{\omega_{\mathcal{G}_X^l}}\},\{{\mathcal{G}_Y^l},{\omega_{\mathcal{G}_Y^l}}\}\Big)\Big).
\end{displaymath}

Finally, by invoking the Isometry theorem and accounting for the direct sum representation~\eqref{direct_sum}, we obtain
  \begin{displaymath}
        W_{\infty} ({D_{\mathbb{G}_{X}}},{D_{\mathbb{G}_{Y}}} ) \leq \max_{1\leq l\leq L} \Big( d_{GH}\Big( \{{\mathcal{G}_{X}^l},  {\omega_{\mathcal{G}_{X}^l}}\},\{{\mathcal{G}_{Y}^l}, {\omega_{\mathcal{G}_{Y}^l}}\}\Big)\Big),
    \end{displaymath}
which concludes the proof.

\subsection*{Experimental Set-up: Computational complexity and sampling of cliques}

The process of tracking clique communities for high values of $k$ is computationally prohibitive, mainly because the clique decision problem is NP-complete and common algorithms require that all the maximal cliques be found first~\cite{CliqueNPComplete:Karp:1972,CliqueHard:Fergal:2012}. 
The aforementioned computational complexity makes it unlikely to apply clique-based methods on the entirety of our large-scale and dense blockchain networks (the Ethereum and Ripple networks have an average of 442788/1192722 and 71337/922084 nodes/edges, respectively). 

To maintain reasonable computation time, we take a subgraph of each network by adopting the maximum weight subgraph approximation method of~\cite{MaxSubgraph:Vassilevska:2006}, which restricts the subgraph size to the nodes of the $p$ most active edges. {Sampling helps us with efficient computation in certain token networks with large sizes. At the same time, sampling nodes has minimal impact on results since most daily token networks have $<$100 nodes. For instance, even for the most traded tokens such as Tronix and Bat, top 150 nodes in daily networks form $75\%$ and $80\%$ of all edges, respectively. The filtered node approach effectively removes $20-25\%$ of edges in calculations, which reduces computation costs.} Hence, with the Ethereum blockchain and Ripple currency networks we only work with 100 and 90 nodes respectively. To mitigate computation efficiency constraints, we set the maximum clique size to $k_{max}=20$ for the Ethereum network and $k_{max}=4$ for Ripple data set, and compute clique community PDs from $k=1$ to $k=k_{max}$ for each network. In addition, an event at time $t$ in the ground truth is marked as detected when a predicted anomaly occurs in $t-2\leq t \leq t+2$; note that trading patterns are variable and may not change the same day as published events in news.

\subsection*{Description of State-Of-The-Art methods} 
\label{subsec:descript}
We select the following state-of-the-art methods for network anomaly detection based on their superior performance proven in most recent experimental studies and their theoretical guarantees~\cite{Yoon2019fast,arlot2019kernel,masuda2019detecting}. The DeltaCon (DC) of~\cite{Koutra2016deltacon} measures the similarity between edgelist pairs, and classifies the events with similarity scores below a specified threshold as anomalous. The two edge monitoring algorithms (EMEu and EMKL) of~\cite{Wang2017fast} account for temporal dependencies by tracking the evolution of the network as a Markov process, thereby focusing on comparison of the probability distribution of edges. The MANE by \cite{li2018multi} calculates the relative difference between pre and post multilayer graph embeddings (i.e., relative difference with Frobenius norm). In turn, gSeg of~\cite{Chen2015scanstat}  is a nonparametric graph-based method that uses two-sample tests based on scan statistics. Among these benchmark methods, the DC and the gSeg are suited for unweighted graph data, while the DC, EMEu and EMKL are applicable for weighted graphs. Since blockchain edges carry transaction amounts, it is natural to use weighted graph based anomaly detection approaches.

\subsection*{Structure of Multiple Hypotheses Testing} 
 Let $\mathcal{A}^l=\{t_1^{l}, t_2^{l},\ldots\, t_{\tau_l}^{l}\}$, $1\leq \tau_l \leq T$, be a set of anomalous events identified in the $T$ sequence of $l$-single layer graphs ${\mathcal{G}}^{l}\in \mathbb{G}$, $1\leq l\leq L$. Then we can test for
\begin{flalign}
\label{H0_weak}
& H_{0}:\:\:
\mathcal{A}^1=\ldots= \mathcal{A}^L=\emptyset
\quad \text{vs} \quad  \nonumber \\
& H_{a}: \:\:\exists\:l \:\:\text{ such that} \:\ \mathcal{A}^l \neq \emptyset, \:\ l \in \{1,\ldots, L\}.
\end{flalign}

The general practice for all statistical hypotheses testing is to control the rate of false positives at a probability known as the level of significance ($\alpha$).
However, the multiple testing approach is significantly prone to varying kinds of false positives rates~\cite{li2013testing}, and the usual kind is the familywise error rate ($FWER$) ~\cite{dudoit2003multiple}, which is defined as:
\[
FWER = P(V \geq 1) = 1-{(1-\alpha_{c})}^{L},
\]
where $V$ is the number of false positives in all $L$ hypotheses, $P$ is probability, $L$ is the number of hypotheses, and $\alpha_{c}$ is the level of significance that determines the rejection of single hypothesis from the multiple set~\cite{dudoit2003multiple,rempala2013permutation}.

We employ a stronger control over $FWER$
by testing
$H_0$ in~(\ref{H0_weak}) against an alternative 
\begin{flalign}
\label{H0_strong}
& H_{0}:\:\:
\mathcal{A}^1=\ldots= \mathcal{A}^L=\emptyset
\quad \text{vs} \nonumber \\
& H_{a}: \:\:\exists\:\: t \in \mathcal{A}^1\cap\ldots \cap \mathcal{A}^L, \quad 1\leq t\leq T.
\end{flalign}

To ensure reliability of conclusions in the multiple testing framework, 
it is essential to control $FWER$ and to require $FWER \leq \alpha$~\cite{dudoit2003multiple}.
  One of the most widely used control procedures is the Bonferroni correction method~\cite{dudoit2007multiple}. Under the Bonferroni correction, we reject the single null hypothesis $H_{0}: \mathcal{A}^{l} = \emptyset$  if and only if $FWER \leq \alpha_{c}$ where $\alpha_{c} = (\alpha/L)$.

\begin{figure*}
\centering
\includegraphics[height = 0.3\textheight,keepaspectratio]{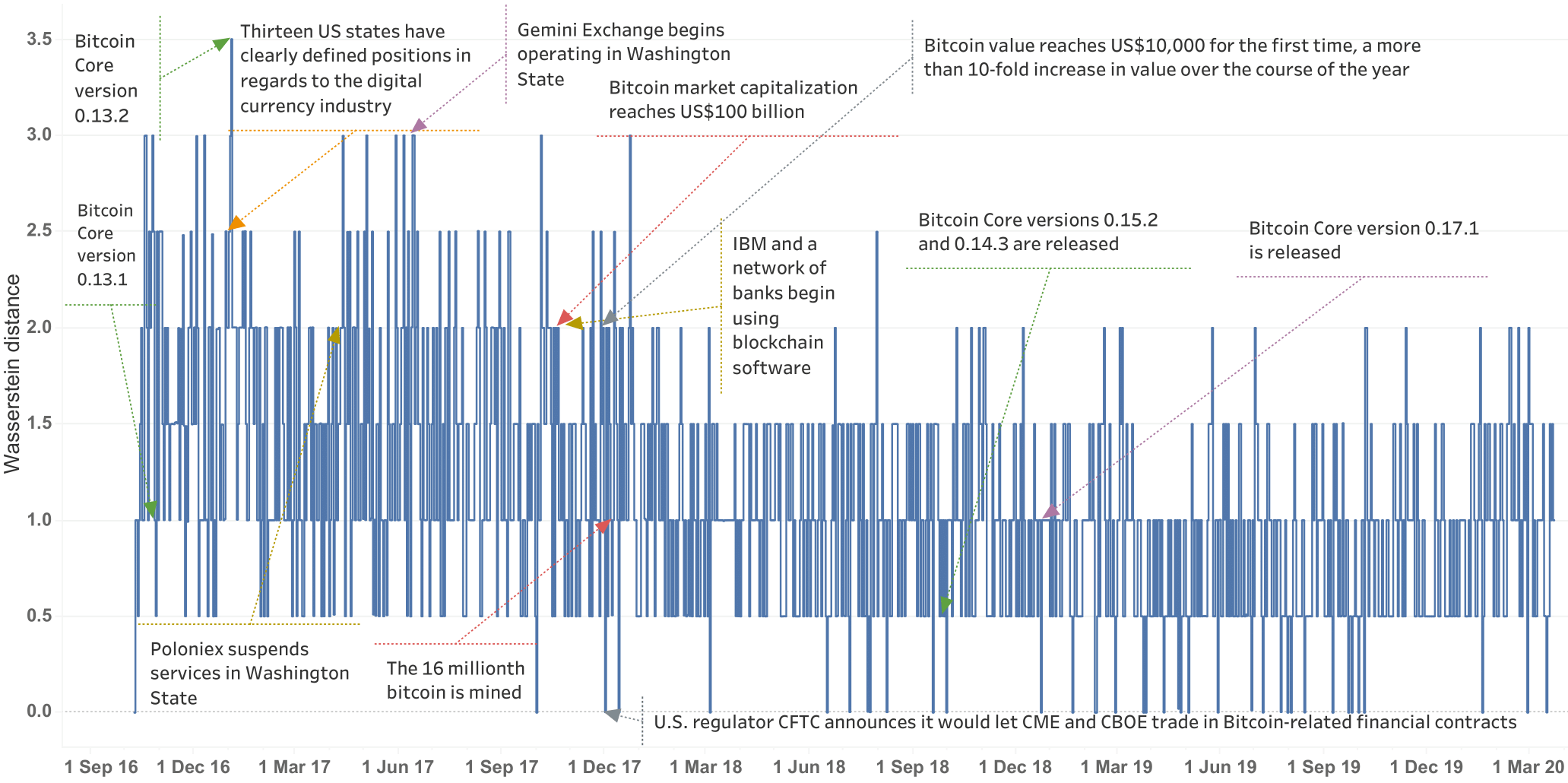}
\caption{Anomalous events detected by \texttt{TAD} for the multilayer Ripple currency network.}
\label{fig:EventMultilayerRipple}
\end{figure*}

\subsection*{Results for the stronger controlled FWER}
At this stage, we test the following multiple hypotheses for the 4 single layer SOTA methds (DC, gSeg, EMEu, EMKL) and our improvised \texttt{S-TAD} method against the \texttt{TAD}:
\begin{flalign}
\label{H0_strong2}
& H_{0}:\:\:
\mathcal{A}^1=\ldots= \mathcal{A}^L=\emptyset
\quad \text{vs} \nonumber \\ \quad
& H_{a}: \:\:\exists\:\: t \in \mathcal{A}^1\cap\ldots \cap \mathcal{A}^L, \quad 1\leq t\leq T.
\end{flalign}

The performance of all methods (\texttt{S-TAD}, DC, gSeg, EMEu, EMKL, \texttt{TAD}) is summarized as true positive (TP), false positive (FP), true negative (TN), false negative (FN), accuracy (Acc.), precision (Prec.) and F1 score values (F1).

\noindent{\bf Results (Ethereum blockchain): } From Tables~\ref{table_Weigth_Ether_FWER} and \ref{table_UnWeigth_Ether_FWER}, we find that \texttt{TAD} delivers the highest detection accuracy (0.954, which is about 5\% greater than what all the other methods deliver) for both the unweighted and weighted networks. 

\begin{table}[!ht]
  \centering
   \caption{Anomaly detection performance for the weighted Ethereum blockchains using strong control of FWER.}
  \begin{tabular}{llllll}
    \hline
     & \texttt{S-TAD}  & DC & EMEu &  EMKL & \texttt{TAD} \\[0.3ex] 
    \hline
     TP   &  0  & 0    & 0    &  0   & 10 \\
     FP    &  0  & 0    & 0    &  0   & 2 \\
     TN   &  138  & 138    & 138  & 138  & 135 \\
     FN   &  14  & 14     & 14   & 14   & 5 \\
     Acc.  & 0.908 & 0.908  & 0.908 & 0.908 & 0.954 \\
    Prec.  & 0.000  & 0.000 & 0.000 & 0.000 & 0.833 \\
    F1  & 0.000 & 0.000  & 0.000 & 0.000 & 0.741 \\
    \hline
    \end{tabular}
     \label{table_Weigth_Ether_FWER}
\end{table}

By comparing the remaining summaries (TP, FP, TN, FN, Prec., F1) to the case we assessed earlier,

we find that with the strong $FWER$ setup, there is no common anomalous event located in the sequence of six layer networks (i.e. the setup of the the six hypotheses makes it harder to locate a common event).  

\begin{table}[!ht]
  \centering
  \caption{Anomaly detection performance for the unweighted Ethereum blockchains using strong control of FWER}
  \begin{tabular}{lllll}
    \hline
     & \texttt{S-TAD}  &  DC  & gSeg & \texttt{TAD} \\
    \hline
     TP   & 0    & 0   & 0   & 10  \\
     FP   & 0    & 0   & 0   & 2 \\
     TN   & 138    & 138   & 138  & 135 \\
     FN   & 14    & 14    & 14   & 5   \\
     Acc. & 0.908  & 0.908 & 0.908 & 0.954 \\
     Prec. & 0.000  & 0.000 & 0.000 & 0.833  \\
     F1 & 0.000  & 0.000 & 0.000 & 0.741 \\
    \hline
    \end{tabular}
      \label{table_UnWeigth_Ether_FWER}
\end{table}

\smallskip

\begin{table}[!ht]
  \centering
  \caption{Anomaly detection performance for the weighted Ripple currency networks using strong control of FWER.}
  \begin{tabular}{llllll}
    \hline
     & \texttt{S-TAD}  & DC & EMEu &  EMKL & \texttt{TAD} \\[0.3ex] 
    \hline
     TP  &  0  & 0    & 0    &  0   & 16 \\
     FP  &  0  & 0    & 0    &  0   & 9 \\
     TN  &  1198  & 1198    & 1198  & 1198  & 1187 \\
     FN  &  64  & 64     & 64   & 64   & 50 \\
     Acc. & 0.949 & 0.949  & 0.949 & 0.949 & 0.953 \\
    Prec.  & 0.000  & 0.000 & 0.000 & 0.000 & 0.640 \\
    F1  & 0.000 & 0.000  & 0.000 & 0.000 & 0.352 \\
    \hline
    \end{tabular}
      \label{table_Weigth_Ripple_FWER}
\end{table}

\noindent{\bf Results (Ripple currency): } Looking at the summaries in Tables~\ref{table_Weigth_Ripple_FWER} and \ref{table_UnWeigth_Ripple_FWER}, we find that the \texttt{TAD} has the highest accuracy (0.953, which is closely matched by the other methods) for the weighted networks. With the unweighted networks, however, the performance of the other methods (0.949) is slightly higher than the accuracy for \texttt{TAD} (0.938). Similar to the Ethereum blockchain results, we notice that the ability of the competing techniques and our \texttt{S-TAD} are strongly disadvantaged when the multiple hypotheses is setup to strongly control $FWER$ (because no unique anomalous event is identified).

\begin{table}[!ht]
  \centering
  \caption{Anomaly detection performance for the unweighted Ripple currency networks using strong control of FWER}
  \begin{tabular}{lllll}
    \hline
     & \texttt{S-TAD}  &  DC  & gSeg & \texttt{TAD} \\
    \hline
     TP   & 0    & 0   & 0   & 11 \\
     FP   & 0    & 0   & 0   & 23 \\
     TN   & 1198    & 1198   & 1198  & 1173 \\
     FN   & 64    & 64    & 64   & 55  \\
     Acc. & 0.949  & 0.949 & 0.949 & 0.938 \\
    Prec. & 0.000  & 0.000 & 0.000 & 0.324 \\
    F1 & 0.000  & 0.000 & 0.000 & 0.220 \\ 
    \hline
    \end{tabular}
      \label{table_UnWeigth_Ripple_FWER}
\end{table}

\subsection*{One-day Example}
\begin{figure*}[!ht]
 	\centering
 	\subfloat[]{\includegraphics[width=0.3\textwidth, keepaspectratio]{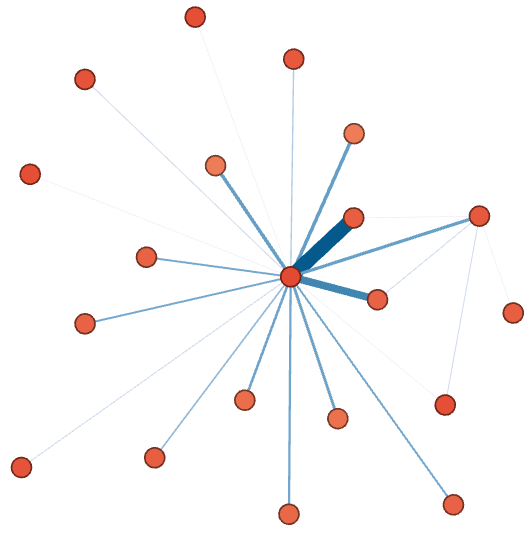}\label{fig:OneDayRippleNet}} \quad
 	\subfloat[]{\includegraphics[width=0.3\textwidth, keepaspectratio]{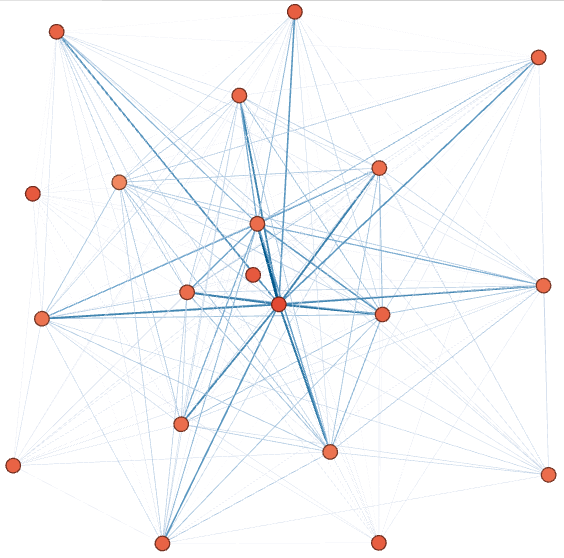}\label{fig:OneDayRippleGD}}\quad
    \subfloat[]{\includegraphics[width=0.3\textwidth, keepaspectratio]{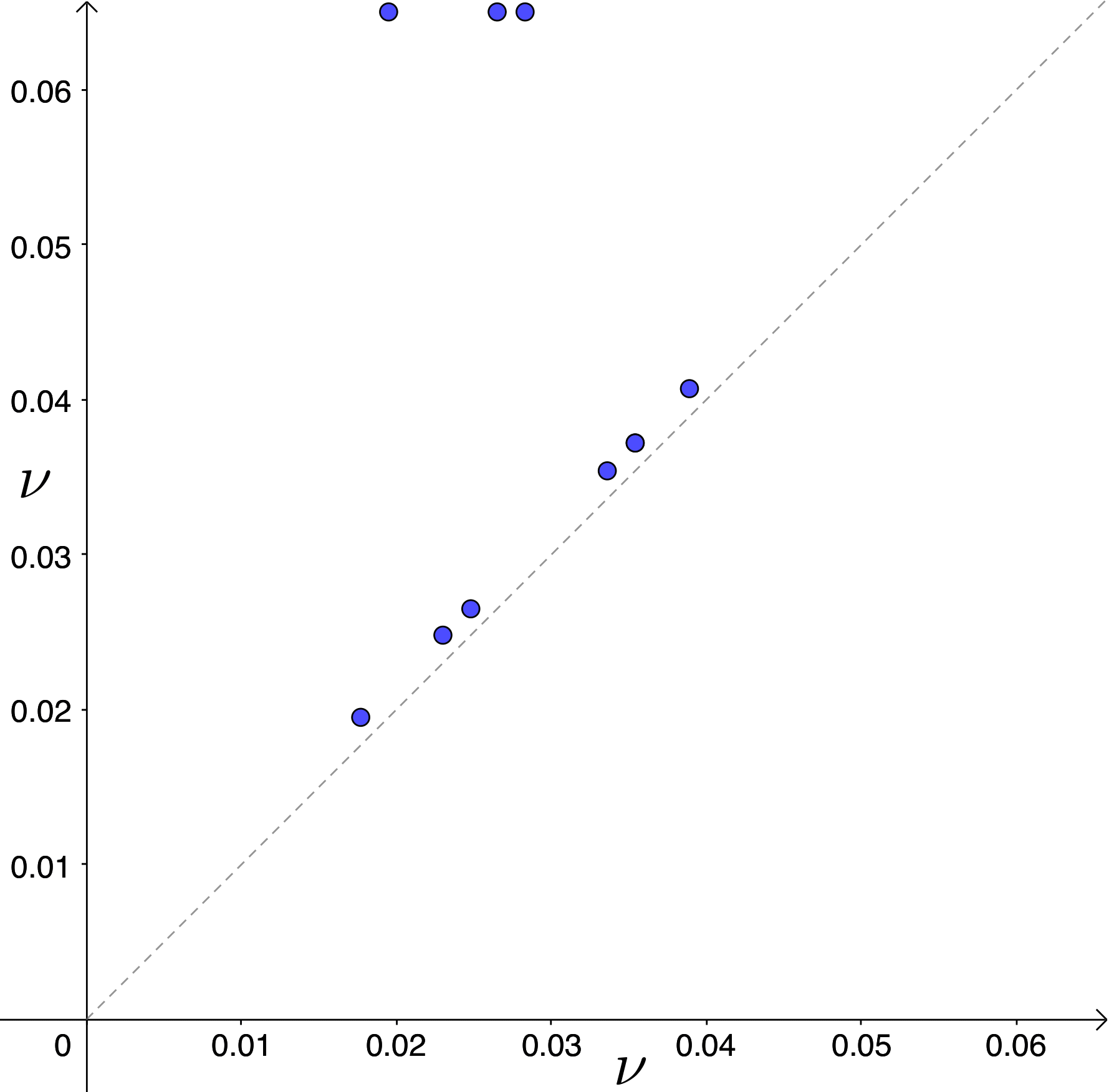}\label{fig:OneDayRipplePD}}
\caption{Processing one-day unilayer graph from Ripple network. (a) One-day graph of CCK currency, twenty connected nodes. (b) Corresponding Geodesic densification. (c) Clique Community Persistence Diagram.}
 \label{fig:NetGDandPD} 
\end{figure*}

\texttt{TAD} computes PDs on geodesic densifications of bloackchain graphs, tracking $k$-clique community persistences based pair-nodes connections. Figure \ref{fig:NetGDandPD} shows a real example on one-day graph of CCK currency, i.e. Ripple currency network. A weighted graph, with twenty connected nodes, is presented in Figure~\ref{fig:OneDayRippleNet}. We compute the geodesic distance between each pair of nodes, i.e. the shortest path length between each node-pair, see Figure
~\ref{fig:OneDayRippleGD}. Finally, significant topological summaries are found
through the clique community persistence procedure, see Figure~\ref{fig:OneDayRipplePD}, using community sizes from $k=1$ to $k=4$. Hence, our methodology preserves  not only direct $k$-cliques communities present in the network, but also communities from indirect connections.

\section*{Acknowledgements}
This work is supported in part by NSF Grants No. ECCS 2039701, DMS 1925346, CNS 1837627, OAC 1828467, IIS 1939728, CNS 2029661 and Canadian NSERC Discovery Grant RGPIN-2020-05665. The authors would like to thank Baris Coskunuzer for insightful discussions.

\bibliographystyle{splncs04}
\bibliography{main.bib}

\end{document}